\documentclass{iopart}
\usepackage{amstext}
\usepackage{amssymb}
\usepackage{graphicx}
\usepackage{pstcol}
\begin{document}

\title{Loop expansion around the Bethe-Peierls approximation for lattice models}

\author{Giorgio Parisi}
\address{Dipartimento di Fisica, Sezione INFN, and Unit\`a INFM,
I Universit\`a di Roma ``La Sapienza'',\\ 
Piazzale Aldo Moro 2, I-00185 Roma, Italy}

\author{Franti\v{s}ek Slanina}
\address{        Institute of Physics,
	Academy of Sciences of the Czech Republic,\\
	Na~Slovance~2, CZ-18221~Praha,
	Czech Republic}
\ead{slanina@fzu.cz}
\date{\today}
\begin{abstract}
We develop an effective field theory for lattice models, in which the
only non-vanishing diagrams exactly reproduce the topology of the lattice.
The Bethe-Peierls approximation appears naturally 
as the saddle point approximation. The corrections to the saddle-point
result can be obtained systematically. We
calculate the lowest loop corrections for magnetisation and
correlation function.

 \end{abstract}

\noindent{\it Keywords\/}: phase transitions and critical phenomena,
mean-field approximations

\pacs{PACS numbers: 05.50.+q 
%Lattice theory and statistics (Ising, Potts, etc.)
}
\maketitle

%\section{Introduction}
There are two ways for doing mean-filed-like approximations in lattice
models. The first one consists in replacing interaction constants decaying
with distance by effective distance-independent interactions. The best
known of these approximation schemes is the Bragg-Williams
approximation \cite{bra_wil_34}. The
second path to a mean-field result 
neglects the correlations induced by the presence of closed
loops on the lattice. This is the Bethe-Peierls (BP) approximation
\cite{bethe_35}.  The
former approach is exact in fully 
connected models, while latter is exact on the Bethe
lattice \cite{baxter_82}. 

The list applications of the BP approximation in statistical physics
is very long. As an example, 
let us mention only the application to spin glasses 
\cite{se_pa_96,walasek_98,mez_par_01,par_tri_02} and
combinatorial optimisation
\cite{mez_par_zec_02}, which attracted much attention recently. Within
this context, the BP approximation essentially coincides with the
cavity method.

The
systematic field-theoretical treatment of the corrections to the
first method is well developed and widely used. On the other hand, to
our knowledge, the
corrections to BP results were not yet formulated in a similar manner.
The actually available approaches rely mostly on cluster variational methods
\cite{kikuchi_51} or expansions in inverse dimension
\cite{geo_yed_91}, where loops of limited size are taken into
account. Recently, a very promising path was paved by Montanari and
Rizzo \cite{mon_riz_05} who were able to consider loops of arbitrary
size, correcting the BP result for Ising model on hypercubic lattice
and also improving the cavity equations for a spin glass.

An important step the direction of field-theory formulation of the BP
approximation was taken by Efetov \cite{efetov_90} in the 
context of
supersymmetric formulation of the Anderson localisation problem. The
method of Efetov consists in writing an effective Lagrangian, with the
property, that the sequence of Feynman diagrams of the corresponding field
theory is equal to
the sum of partition function of the original model, plus partition
function of  the same model wrapped several times around.  It is
expected that in
thermodynamic limit the wrapping is unessential.

Then, the saddle-point
equations for the effective field theory correspond exactly to the BP
approximation. In this way, Efetov was able to locate the localisation
transition and calculate the correlation function to the zero-loop
order. It is worth noting, however, that the zero-loop order on the
correlation function corresponds already to the lowest correction to
the BP result, in which the connected correlation function is exactly
zero. 

The purpose of our present work is to show, how the Efetov's method
can be applied to general lattice model, especially magnetic models
like Ising or Ginsburg-Landau models. We show, how the well-known BP
results are generated within the formalism and calculate the
lowest-order corrections, which are one-loop for the magnetisation and
zero-loop for the correlation function.

%\section{General formalism on arbitrary graph}

%\subsection{Lattice, state variables}
We shall consider a system defined on a lattice represented by 
 an oriented graph
$\{\mathcal{N},\mathcal{E}\}$, where $\mathcal{N}$ is the set of
nodes, while $\mathcal{E}\subset \mathcal{N}\times\mathcal{N} $ is the
set of oriented edges. 
For each node $r\in\mathcal{N}$ define
$\mathcal{I}_r=\{a\in\mathcal{E} |\,\exists r'\in\mathcal{N}:a=(r',r)\}$ and
$\mathcal{O}_r=\{a\in\mathcal{E} |\,\exists r'\in\mathcal{N}:a=(r,r')\}$
the sets of incoming and outgoing edges, respectively. 
For further convenience we denote
$\underline{a}$ the initial and $\bar{a}$ the final node of the
edge $a\in\mathcal{E}$, i. e. $a=(\underline{a},\bar{a})$.

Each node $r$ is described by the state variable
$\phi_r\in\mathcal{S}$. We shall treat both discrete and continuous
$\phi$ on equal footing, denoting by $\sum_\phi$ summation in
discrete case and integration in continuous case. The state of the
system is described by $\Phi\in\mathcal{S}^\mathcal{N}$.
For the Efetov method to work it is essential to decompose the
Hamiltonian of the system into node and edge parts,
\begin{eqnarray}
&H=&H_\mathcal{N}+H_\mathcal{E}
\nonumber\\
&H_\mathcal{N}=&-\sum_{r\in\mathcal{N}}V_r(\phi_r)\\
&H_\mathcal{E}=&
-\sum_{r\in\mathcal{E}}U(\phi_{\underline{a}},\phi_{\bar{a}})\; .
\nonumber
\end{eqnarray}
Therefore, all the terms in the partition function 
are products of node and edge factors,
\begin{equation}
Z=\sum_\Phi\mathrm{e}^{-H[\Phi]}=
\sum_\Phi
\left(\prod_{a\in\mathcal{E}}\gamma(\phi_{\underline{a}},\phi_{\bar{a}})\right)
\left(\prod_{r\in\mathcal{N}}\zeta_r(\phi_r)\right)
\end{equation}
where $\gamma(\phi,\phi')=\mathrm{e}^{U(\phi,\phi')}$
and $\zeta_r(\phi)=\mathrm{e}^{V_r(\phi)}$.
When appropriate we will use also matrix notation 
$\zeta_r(\phi,\phi')=\zeta_r(\phi)\delta(\phi-\phi')$. We will also
use the inverse matrix $\gamma^{-1}$, where obviously
$\sum_{\phi''}\gamma(\phi,\phi'')\gamma^{-1}(\phi'',\phi')=\delta(\phi-\phi')$

In order to build the effective field theory, 
we introduce complex auxiliary fields for each
edge $a\in\mathcal{E}$ and each value of the physical field 
$\phi\in\mathcal{S}$, 
$x_a(\phi)\in\mathbb{C}$ and its complex conjugate $\bar{x}_a(\phi)$.

The Efetov Lagrangian is similarly decomposed into edge and node part
\begin{equation}
\mathcal{L}=\mathcal{L}_0+\mathcal{L}_1
\label{eq:lagrangian}
\end{equation}
where the edge part is simply
\begin{equation}
\mathcal{L}_0=\sum_{a\in\mathcal{E}}\sum_{\phi\phi'}
\bar{x}_a(\phi)\gamma^{-1}(\phi,\phi')x_a(\phi')
\end{equation}
while the node part is slightly more complicated
\begin{equation}
\mathcal{L}_1=-\sum_{r\in\mathcal{N}}\sum_{\phi}
\zeta_r(\phi)
\left(\prod_{a\in\mathcal{I}_r}x_a(\phi)\right)
\left(\prod_{a\in\mathcal{O}_r}\bar{x}_a(\phi)\right)\; .
\end{equation}

The key quantity is 
the Efetov partition
function
\begin{equation}
\mathcal{Z}=\frac{%
\int\mathrm{d}[x,\bar{x}]\mathrm{e}^{-\mathcal{L}[x,\bar{x}]}
}{%
\int\mathrm{d}[x,\bar{x}]\mathrm{e}^{-\mathcal{L}_0[x,\bar{x}]}
}
=\langle\mathrm{e}^{-\mathcal{L}_1}\rangle_0
\label{eq:efetovpartitionfunction}
\end{equation}
where we denoted the average with respect to $\mathcal{L}_0$ 
as $\langle\;\rangle_0$.

The diagrams contributing to (\ref{eq:efetovpartitionfunction})
correspond to the lattice of the original system, wrapped $m$-times 
around, where $m=0,1,2,3,...$. Indeed, the first contribution is
obviously 1 and in all the remaining diagrams the lines are exactly
the edges of the lattice and the graph vertices are the nodes. Each
node and every edge must appear the same number (which is $m$) of times, but
they may be connected by various different manners. That is how the
wrappings come around. 
The linked-cluster theorem enables us to write
\begin{equation}
\mathcal{Z}=\mathrm{e}^{\mathcal{Z}_c}
\end{equation}
where $\mathcal{Z}_c$ is the sum of all connected components coming
from the wrapping. The first term is $Z$, second term contains various
possibilities to wrap system twice without leaving two disconnected
parts etc.

The averages of physical quantities, like magnetisation and
correlation function, are calculated by making derivatives of
$\mathcal{Z}$ with respect to $\zeta_r(\phi)$.
Using the notation
$\partial_r(\phi)\equiv\zeta_r(\phi)\frac{\partial}{\partial\zeta_r(\phi)}$
we find that the magnetisation on site $r$ is
\begin{equation}
M_r=
\frac{%
\sum_\phi \phi\,\partial_r(\phi)\mathcal{Z}
}{%
\sum_\phi\partial_r(\phi)\mathcal{Z}}
\label{eq:magnetization}
\end{equation}
and the 
connected correlation function $C(r,r')$ 
can be obtained by derivative with respect to local magnetic field, giving
\begin{eqnarray}
&C(r,r')=&
\frac{%
\sum_{\phi\phi'} \phi\,\phi'\,\partial_r(\phi)\partial_{r'}(\phi')\mathcal{Z}
}{%
\sum_\phi\partial_r(\phi)\mathcal{Z}}
-\\
&&-\frac{%
\sum_{\phi} \phi\,\partial_{r}(\phi)\mathcal{Z}
}{%
\sum_\phi\partial_r(\phi)\mathcal{Z}}
\frac{%
\sum_{\phi\phi'} \phi'\,\partial_r(\phi)\partial_{r'}(\phi')\mathcal{Z}
}{%
\sum_\phi\partial_r(\phi)\mathcal{Z}}\; .
\nonumber
\end{eqnarray}

Technically, the treatment is complicated by very high symmetry of the
Lagrangian (\ref{eq:lagrangian}), which cannot be easily
parametrised. 
Let
$\eta_a=\mathrm{e}^{\mathrm{i}\epsilon_a}\in U(1)$ for each edge
$a\in\mathcal{E}$. The field variables transform as 
\begin{eqnarray}
&x_a(\phi)\to&\eta_a x_a(\phi)\\
%& &\\
&\bar{x}_a(\phi)\to&\bar{\eta}_a \bar{x}_a(\phi)
\end{eqnarray}
with one condition at each vertex
$\big(\prod_{a\in\mathcal{I}_r}\eta_a\big)
\big(\prod_{a\in\mathcal{O}_r}\bar{\eta}_a\big)=1$ which can be
interpreted as the requirement of current conservation .
 The number of free parameters of
the symmetry group is $|\mathcal{E}|-|\mathcal{N}|$,
i. e. proportional to the system size.

As a zeroth approximation we can calculate the observable quantities using 
saddle-point method. This amounts to solving the coupled set of equations
\begin{equation}
\frac{\partial\mathcal{L}[x^{(s)},\bar{x}^{(s)}]}{\partial x_a(\phi)}=
\frac{\partial\mathcal{L}[x^{(s)},\bar{x}^{(s)}]}{\partial \bar{x}_a(\phi)}=0
\end{equation}
which means
\begin{eqnarray}
&x_a^{(s)}(\phi)&=\sum_{\phi'}\gamma(\phi,\phi')
\prod_{b\in\mathcal{I}_{\underline{a}}}x_b^{(s)}(\phi')
\prod_{{b\in\mathcal{O}_{\underline{a}}\, 
(b\ne a)}}\bar{x}_b^{(s)}(\phi')
\nonumber\\
%& &\\
&\bar{x}_a^{(s)}(\phi)&=\sum_{\phi'}\gamma(\phi,\phi')
\prod_{{b\in\mathcal{I}_{\overline{a}}\, 
(b\ne a)}}x_b^{(s)}(\phi')
\prod_{b\in\mathcal{O}_{\overline{a}}}\bar{x}_b^{(s)}(\phi')\; .
\nonumber
\end{eqnarray}
This set of equations has infinitely many equivalent solutions 
 related by the symmetry group operations. In the following, we shall
 always select the real uniform solution
 $x_a^{(s)}=\bar{x}_a^{(s)}=x\,\forall a\in\mathcal{E}$.

To take a concrete example for the geometry, let us consider the
$d$-dimensional hypercubic lattice. 
%
%The node and edge contributions to
%the Hamiltonian are as usual $V_r(\phi)=h\,\phi$ and
%$U(\phi,\phi')=J\,\phi\,\phi'$.  
We will use indices $\alpha,\beta,...\in\{1,2,...,d\}$ to denote one of the 
Cartesian axes and indices $\sigma,\theta,...\in\{-1,1\}$ to
indicate one if the two orientations along one of the $d$ axes. The
$2d$ possible directions will be denoted by indices
$\mu,\nu,...\in\{1,...,2d\}$. On hypercubic lattice it is convenient
to change the notation for the field variables as
$\tilde{x}_{r\,2\alpha-1}(\phi)=x_{(r-e_\alpha ,r)}(\phi)$ and
$\tilde{x}_{r\,2\alpha}(\phi)=\bar{x}_{(r,r+e_\alpha )}(\phi)$, where
$e_\alpha$ are the unit translation vectors along the Cartesian axes.

The real uniform saddle point is the solution of $|\mathcal{S}|$
coupled equations
\begin{equation}
x(\phi)=\sum_{\phi'}\gamma(\phi,\phi')\zeta(\phi')x^{2d-1}(\phi')\; .
\label{eq:forx}
\end{equation}
If we solve them, we can calculate the magnetisation
\begin{equation}
M=
\frac{%
\sum_\phi \phi \zeta(\phi)x^{2d}(\phi)
}{%
\sum_\phi \zeta(\phi)x^{2d}(\phi)
}
\end{equation}
while the connected correlation function is identically zero.
These results coincide with the Bethe-Peierls approximation, or,
equivalently, with the solution of the corresponding model on 
Bethe lattice with coordination number $2d$
 \cite{baxter_82}.

Fluctuations around the saddle point are denoted as
$y_{r\mu}(\phi)=\tilde{x}_{r\mu}(\phi)-x(\phi)$ and the Lagrangian is
\begin{equation}
\mathcal{L}=\mathcal{L}[x^{(s)},\bar{x}^{(s)}]
+\frac{1}{2}\sum_{rr'}\sum_{\mu\nu}\sum_{\phi\phi'}
y_{r\mu}(\phi)\,M_{r\mu,r'\nu}(\phi,\phi')\,y_{r'\nu}(\phi')+\ldots
\end{equation}
The Fourier transform of the 
bare propagator $G=M^{-1}$ can be written explicitly
\begin{eqnarray}
&&\hat{G}_{\mu\nu}(p|\phi,\phi')=
\label{eq:barepropagator}
\\
&&=\sum_\lambda
x^{1-d}(\phi)\zeta^{-1/2}(\phi)B(\phi,\lambda)
\hat{g}_{\mu\nu}(p|\lambda)
B(\phi',\lambda)\zeta^{-1/2}(\phi')x^{1-d}(\phi')
\nonumber
\end{eqnarray}

We must spend some time explaining the
 factors entering the expression (\ref{eq:barepropagator}). 
The matrix function
%
%\begin{equation}
%\hat{g}(p|\omega)=[\omega^{-1}m_1(p)-m_0]^{-1}
%\end{equation}
%
$\hat{g}(p|\omega)=[\omega^{-1}m_1(p)-m_0]^{-1}$
with $m_{0\mu\nu}=1-\delta_{\mu\nu}$ and $m_{1}(p)_{\mu\nu}=
\delta_{\alpha\beta}(1-\delta_{\sigma\theta})$
depends only on the geometry of the lattice,
but not on the nature of the spin variables $\phi$. Straightforward
algebra gives
\begin{eqnarray}
&\hat{g}_{\mu\nu}(p|\omega)=&\frac{1}{1-\omega^2}\bigg[
\delta_{\alpha\beta}(1-\delta_{\sigma\theta})\,
\mathrm{e}^{\mathrm{i}p_\mu}\,\omega
-
\nonumber\\[-2mm]
& &
\label{eq:smallg}
\\[-2mm]
&&
-\delta_{\mu\nu}\,\omega^2
+\frac{\omega^2\,(\mathrm{e}^{\mathrm{i}p_\mu}-\omega)
(\mathrm{e}^{\mathrm{-i}p_\nu}-\omega)
}{
1-2c(p)\,\omega+(2d-1)\,\omega^2}
\bigg]
\nonumber
\end{eqnarray}
where $c(p)=\sum_{\alpha=1}^d\cos p_\alpha$.

The numbers $\lambda$ form the set of eigenvalues of the (symmetric)
matrix
\begin{equation}
\widetilde{\gamma}(\phi,\phi')=x^{d-1}(\phi)\zeta^{1/2}(\phi)
\gamma(\phi,\phi')\zeta^{1/2}(\phi')x^{d-1}(\phi')
\end{equation}
and $B(\phi,\lambda)$ are the corresponding normalised
eigenvectors. Contrary to the matrix $\hat{g}(p|\omega)$, the
eigenvalues $\lambda$ and the eigenvectors bear the information on the
nature of the variables $\phi$, but conversely the dependence on the
lattice structure is only weak. In fact, the lattice enters only
through $x(\phi)$, which for transitionally invariant lattices depends
only on the coordination number, as can be seen from the saddle-point
equations (\ref{eq:forx}). In such a way the Efetov formalism decouples
to large extent the effects of the lattice structure from the effects
due to specific choice of the Hamiltonian living on the lattice. 

It is easy to see that one of the eigenvalues of 
$\widetilde{\gamma}$
is always equal to $1$. Indeed, from the saddle-point equation
(\ref{eq:forx}), multiplying both sides by
$x^{d-1}(\phi)\zeta^{1/2}(\phi)$ we conclude that
 the vector
\begin{equation}
v_1(\phi)=x^d(\phi)\zeta^{1/2}(\phi)
\end{equation}
is eigenvector of $\widetilde{\gamma}$ and the corresponding eigenvalue is
$1$. This brings about a serious problem in the calculation of the
diagrams, because the bare propagator $\widehat{g}(p|\omega)$ diverges
for $\omega\to 1$, as can be seen in Eq. (\ref{eq:smallg}). The source
of the divergence is the continuous degeneracy of the saddle point,
as mentioned above. Usually such situation is solved by explicit
integration over the degenerate saddle points, but here it is
impractical, because the symmetry group cannot be easily
parametrised. Fortunately enough, the physical observable quantities
contain such combinations of the propagators that the divergences
stemming from various terms
cancel each other. We verified this property by explicit calculations
of the magnetisation and correlation function up to one-loop level. 

In the diagrammatic expansions all summations over field variables
$\phi$ are reduced to sums over the set of eigenvalues $\lambda$, 
so the lines correspond to the propagators
$g(p|\lambda)$. The vertices are of two kinds. As a small filled
circle we depict a vertex involving mere summation over the field
variable, i. e. 
$
\begin{pspicture}(-0.1,-0.08)(0.1,0.2)
\psline{*-*}(0,0)(0,0)
\end{pspicture}
=\sum_\phi\ldots$, 
where the empty circle means summation of terms containing factor
$\phi$, i. e.  
$
\begin{pspicture}(-0.1,-0.08)(0.1,0.2)
\psline{o-o}(0,0)(0,0)
\end{pspicture}
=
\sum_\phi\,\phi\ldots$. Later, we shall see also a vertex denoted by
empty square, denoting summation with factor $\phi^2$. Generally, we
denote the
$n$-legged vertices as
\begin{eqnarray}
\begin{pspicture}(-0.1,-0.08)(0.1,0.2)
\psline{*-*}(0,0)(0,0)
\end{pspicture}
&=&
\Vert v_1 \Vert^{2-n}D_k(\lambda_1,\lambda_2,\ldots,\lambda_n)=\\
&=&\Vert v_1 \Vert^{2-n}\sum_\phi B^{2-n}(\phi,1)
\prod_{l=1}^n B(\phi,\lambda_l)\; .
\nonumber
\end{eqnarray}
The analogous  
cases of empty circles and squares are distinguished by writing one or
two overlines over $D$, respectively. 

Having all elements at hand, we proceed to computation of the one-loop
correction to the magnetisation.
We obtain pictorially
\begin{equation}
M
=\frac{%
\begin{pspicture}(-0.1,0)(0.1,0.2)
\psline{o-o}(0,0)(0,0)
\end{pspicture}
}{%
\begin{pspicture}(-0.1,0)(0.1,0.2)
\psline{*-*}(0,0)(0,0)
\end{pspicture}
}+
\frac{1}{
\begin{pspicture}(-0.1,0)(0.3,0.2)
\psline{*-*}(0,0)(0,0)
\psline{*-*}(0.2,0)(0.2,0)
\end{pspicture}
}
\bigg[
\left(
\begin{pspicture}(-0.3,0)(0.5,0.6)
\pscircle(0,0.3){0.3}
\psline{o-o}(0,0)(0,0)
\psline{*-*}(0.4,0)(0.4,0)
\end{pspicture}
-
\begin{pspicture}(-0.3,0)(0.5,0.6)
\pscircle(0,0.3){0.3}
\psline{*-*}(0,0)(0,0)
\psline{o-o}(0.4,0)(0.4,0)
\end{pspicture}
\right)
+
\left(
\begin{pspicture}(-0.4,0.3)(0.5,1)
\pscircle(0,0.7){0.3}
\psline{o-*}(0,0)(0,0.4)
\psline{*-*}(0.4,0)(0.4,0)
\end{pspicture}
-
\begin{pspicture}(-0.4,0.3)(0.5,1)
\pscircle(0,0.7){0.3}
\psline{*-*}(0,0)(0,0.4)
\psline{o-o}(0.4,0)(0.4,0)
\end{pspicture}
\right)
\bigg]
+\ldots\; .
\label{eq:magnetizationdiagoneloop}
\end{equation}
Writing explicitly the value of the diagrams we can see, that the formal
expansion parameter is $\Vert v_1\Vert ^{-2}$. We should note,
however, that this parameter is not actually small; its value is at
least equal $1$.  
Collecting all the terms to the order
$\Vert v_1\Vert ^{-2}$ we have finally
\begin{eqnarray}
M&=&\overline{D}_0+\frac{1}{2\Vert v_1\Vert ^{2}}
\sum_{\lambda(\ne 1)}\Big[
(\overline{D}_2(\lambda,\lambda)-\overline{D}_0)
+
\nonumber
\\
&&+\left(1-\frac{1}{d}\right)\sum_{\lambda' (\ne 1)}
\overline{D}_1(\lambda')D_3(\lambda',\lambda,\lambda)
\hat{g}_{..}(0|\lambda')
\Big]g_A(0|\lambda)
\label{eq:magnetisation}
\\
&&+O(\Vert v_1\Vert ^{-4})
\nonumber
\end{eqnarray}
The combinations of the propagators occurring in the formula are
\begin{equation}
\hat{g}_{..}(0|\lambda')
\equiv\sum_{\mu\nu}\hat{g}_{\mu\nu}(0|\lambda')
=\frac{2d\lambda'}{1-(2d-1)\lambda'}
\end{equation}
and
\begin{eqnarray}
g_A(0|\lambda)&\equiv&\sum_{\mu\nu}(1-\delta_{\mu\nu})g_{\mu\nu}(0|\lambda)=
\frac{
1-(2d-1)\lambda^2
}{
1-\lambda
}
\Bigg[
-\frac{1}{1+\lambda}
+
\\
&&+
\int_{-\pi}^\pi
\frac{\mathrm{d}^d p}{(2\pi)^d} \,
\frac{1}{
1-(2d-1)\lambda+\frac{2\lambda}{1-\lambda}\sum_\alpha^d(1-\cos p_\alpha)
}\Bigg]
\end{eqnarray}
We can see that the phase transition is located at
$\lambda=\lambda_c\equiv 1/(2d-1)$, as in the BP approximation. 
To calculate the correction to the transition temperature we would
need to make partial resummation of the series, which is feasible, but
the results will be presented in subsequent publication. Instead, we
shall estimate the correction to the transition temperature by a
simple argument later.

As a special case, let us consider the Ising model.  
The matrix $\widetilde{\gamma}$ has only two
eigenvalues, one of which is always equal $1$, as already stated. For
now, denote $\lambda$ the other one.
There is only single term in both summations of
(\ref{eq:magnetisation}) and we have (superscript $(0,1)$ indicates that we
include only zero and one-loop contributions)
\begin{equation}
M^{(0,1)} =(\cos 2\vartheta)\Bigg[1-\frac{1}{2\Vert v_1\Vert^2}
\left(1+\left(1-\frac{1}{d}\right)\hat{g}_{..}(0|\lambda)\right)g_A(0|\lambda)
\Bigg]
\label{eq:magnetisationising}
\end{equation}
where the angle $\vartheta$ is related by equation
$v_1=\Vert v_1\Vert \left(\begin{array}{l}\cos\vartheta\\ 
\sin\vartheta \end{array}\right) $ to the eigenvector of $\tilde{\gamma}$
corresponding to eigenvalue 1. This expression suggests that the
applicability of the expansion relies on the smallness of the
effective expansion parameter $g_A/\Vert v_1\Vert^2$.

Far from the transition point we have
$\lambda\ll\lambda_c$. Indeed, for small $h$ we obtain in the
high-temperature phase $\lambda\simeq J$, while in the low-temperature
phase $\lambda\simeq \mathrm{e}^{-4J(d-1)}$, so we can expand the result in powers of
$\lambda$. 
Keeping only the lowest terms, i. e. up to order
$O(\lambda^5)$, we get  
\begin{equation}
M\simeq (\cos 2\vartheta)\Big[1-\frac{1}{\Vert v_1\Vert^2}
\left(1+2(d-1)\lambda\right)4d(d-1)\lambda^4 \Big]\; .
\end{equation}

In the saddle-point approximation the connected correlation function
vanishes. However, the first correction is found already at zero-loop
level. To be consistent with the calculation of the magnetisation, 
we proceed by differentiating the magnetisation with respect to local
magnetic field. We find, diagrammatically (superscript $0$ indicating
that we keep only zero-loop terms) 
\begin{eqnarray}
C^{(0)}(r-r')&=&\frac{\partial}{\partial h_{r'}}M_r^{(0)}
=
\nonumber
\\[-2mm]
&&\label{eq:correlfzeroloop}
\\[-2mm]
&=&\frac{1}{%
\begin{pspicture}(0,0)(0.4,0.2)
\psline{*-*}(0.1,0.1)(0.1,0.1)
\psline{*-*}(0.3,0.1)(0.3,0.1)
\end{pspicture}
}
\big[\left(
\begin{pspicture}(-0.1,0)(1,0.1)
\psline{*-*}(0,0)(0,0)
\psline{o-o}(0.3,0)(0.9,0)
\end{pspicture} 
-
\begin{pspicture}(-0.1,0)(1,0.1)
\psline{o-o}(0,0)(0,0)
\psline{*-o}(0.3,0)(0.9,0)
\end{pspicture} 
\right)
+\left(
\begin{pspicture}(-0.2,0)(0.4,0.2)
\psframe(-0.07,-0.07)(0.1,0.1)
\psline{*-*}(0.3,0)(0.3,0)
\end{pspicture}
-
\begin{pspicture}(0,0)(0.4,0.2)
\psline{o-o}(0.1,0)(0.1,0)
\psline{o-o}(0.3,0)(0.3,0)
\end{pspicture}
\right)\delta(r-r')\big]
\nonumber
\end{eqnarray}
and therefore
\begin{eqnarray}
C^{(0)}(r)&=&\delta(r)\bigg[\overline{\overline{D}}_0-\overline{D}_0^2
-\sum_{\lambda(\ne 1)}\overline{D}_1^2(\lambda)
  \bigg]
+\sum_{\lambda(\ne 1)}\overline{D}_1^2(\lambda)\times
\nonumber\\[-2mm]
&&\\[-2mm]
&\times&
\int_{-\pi}^\pi
\frac{\mathrm{d}^d p}{(2\pi)^d} \,
\frac{\mathrm{e}^{\mathrm{i}pr}(1+\lambda)}{
1-(2d-1)\lambda+\frac{2\lambda}{1-\lambda}\sum_\alpha^d(1-\cos p_\alpha)
}\; .
\nonumber
\end{eqnarray}
Note that derivative with respect to magnetic field produces also
on-site terms (proportional to $\delta(r-r')$) which contain summing
over $\phi$ a function multiplied once more by $\phi$. From here the
new vertex 
$\begin{pspicture}(-0.2,0)(0.2,0.2)
\psframe(-0.07,-0.07)(0.1,0.1)
\end{pspicture}
=\overline{\overline{D}}_0=\sum_\phi \phi^2
B^2(\phi,1)$.

For Ising model in high temperature phase at zero magnetic field
$\overline{D}_1(\lambda)=-1$, $\overline{D}_0=0$,
$\overline{\overline{D}}_0=D_0=1$ and we have
\begin{equation}
C^{(0)}(r)=\int_{-\pi}^\pi
\frac{\mathrm{d}^d p}{(2\pi)^d} \,
\frac{\mathrm{e}^{\mathrm{i}pr}(1+\lambda)}{
1-(2d-1)\lambda+\frac{2\lambda}{1-\lambda}\sum_\alpha^d(1-\cos p_\alpha)
}
\end{equation}
We were able to compute also the one-loop contributions to the
connected correlation function, verifying explicitly that the
divergences stemming from the propagators $g_{\mu\nu}(p,\omega)$,
$\omega\to 1$ exactly cancel, leaving a finite result. The final
formulae are rather lengthy, though, and will be reported elsewhere
\cite{par_sla_preparation}. 

Finally, let us consider the shift in the transition temperature for
the Ising model. A
consistent calculation would require explicit partial resummation of
the loop expansion, as noted before. However, we can get an estimate
of the correction by the following simple consideration. The
magnetisation (\ref{eq:magnetisationising}) has the form
$M^{(0,1)}=M^{(0)}(h)\,[1-m^{(1)}(h)]$, where $M^{(0)}$ is the BP (or
saddle-point) value for the magnetisation 
and $m^{(1)}(h)$ is an even function of
$h$, so $\lim_{h\to 0}\frac{\mathrm{d}}{\mathrm{d}h}m^{(1)}(h)=0$. This
follows from the fact that both $\lambda$ and $\Vert v_1\Vert^2$ are
even functions of $h$. 

The expression (\ref{eq:magnetisationising}) can serve as a starting
point for calculating first correction in the loop expansion for the
inverse susceptibility $\chi^{-1}=\lim_{M\to
0}\frac{\mathrm{d}h}{\mathrm{d}M}$. So, up to order $\Vert
v_1\Vert^{-2}$ we have
\begin{equation}
(\chi^{-1})^{(0,1)}=(\chi^{-1})^{(0)}[1+m^{(1)}(0)]
\end{equation}
and the shifted transition temperature is given by the condition
$m^{(1)}(0)=-1$. We get the following equation for $\lambda_c=\tanh
J_c$
\begin{equation}
\lambda_c=\frac{1}{2d-1}
+\frac{2(d-1)}{2d-1}\frac{\lambda_c\,g_A(0|\lambda_c)
}{
2^{-\frac{1}{d-1}}\,(1-\lambda_c^2)^\frac{d}{2d-2}
+g_A(0|\lambda_c) 
}\; .
\label{eq:critpoint}
\end{equation}
This equation can be solved by iteration, the first step consisting in
inserting the BP value $\lambda_c^{(0)}=(2d-1)^{-1}$ to the right-hand
side. This way we get the following estimates for the corrected
transition temperature (the numerical integration was performed in
Maple): 
%
%\begin{eqnarray}
%&J_c=0.2372651\ldots&\text{ for }d=3\nonumber\\
%&J_c=0.1520003\ldots&\text{ for }d=4\label{eq:critpointnumer}\\
%&J_c=0.1144870\ldots&\text{ for }d=5\nonumber\\
%\end{eqnarray}
%
$J_c=0.2372651\ldots$ for $d=3$, $J_c=0.1520003\ldots$ for $d=4$, and
$J_c=0.1144870\ldots$ for $d=5$. 
These values are strikingly close to those calculated in
Ref. \cite{mon_riz_05}. Moreover, if we linearise the RHS of
(\ref{eq:critpoint}) in the supposedly small value of
$g_A(0|\lambda_c)$, we get nearly identical expression for the
correction as reported in  \cite{mon_riz_05} (equation (3.17)), 
the difference being
only in a numerical factor which goes to $1$ for $d\to\infty$. 
It suggests that our approach and that of Ref.  \cite{mon_riz_05} are
intimately related. The details if that relation are not yet clear, though.

To conclude, we developed a systematic procedure how to calculate
corrections to Bethe-Peierls approximation, which can be used for any
equilibrium statistical model on lattice. We showed it 
in the simplest case of the Ising model on hypercubic lattice,
obtaining finally corrections to the transition temperature.
In principle, our method should work on lattices of arbitrary
geometry, but the practical calculation of the
diagrams relies on translational symmetry. An application to
topologically disordered systems would be possible, but it would
require
 some additional
approximations.

\section*{Acknowledgements}
F.S. wishes to thank to INFN section of University of Rome 
``La Sapienza'' for financial support and kind hospitality. The work
was supported by the GAAV, project IAA1010307.

%\pagebreak
\end{document}